\documentclass[article]{aa}

\usepackage{txfonts}
\usepackage{multirow}
\usepackage{verbatim}
\usepackage[usenames,dvipsnames]{color}
\usepackage{sidecap}
\usepackage{natbib}
\usepackage[dvips]{graphicx}
\RequirePackage{graphicx}
\RequirePackage{epsfig}
\RequirePackage{enumerate}
\RequirePackage{tabularx}
\RequirePackage{amssymb}
\RequirePackage{rotating}
\usepackage{lscape}

\def\lsim{\mathrel{\rlap{\lower4pt\hbox{$\sim$}}
    \raise1pt\hbox{$<$}}}                
\def\gsim{\mathrel{\rlap{\lower4pt\hbox{$\sim$}}
    \raise1pt\hbox{$>$}}}                

\def\src{IRAS~4A}
\def\asec{\rlap.{''}}
\def\deg{$^\circ$}

\begin{document}

\title{
Comparing star formation models with interferometric observations 
of the protostar NGC~1333~IRAS~4A. I. \\ 
Magnetohydrodynamic collapse models
\thanks{Based on observations carried out with the SMA telescope.}
}

\author{
Pau Frau \inst{1}
\and
Daniele Galli \inst{2}
\and
Josep M.\ Girart \inst{1}
} 

\offprints{
P.\ Frau,
\email{frau@ice.cat}
}

\institute{
Institut de Ci\`encies de l'Espai (CSIC-IEEC), Campus UAB, Facultat de Ci\`encies, Torre C-5p, 08193 Bellaterra, Catalunya, Spain
\and
INAF-Osservatorio Astrofisico di Arcetri, Largo E.~Fermi 5, 50125 Firenze, Italy
}

\date{
Received ???
/ 
Accepted ???
}

\titlerunning{ Comparing star formation models with interferometric
observations. I: MHD collapse models}

\authorrunning{}

\abstract
{Observations of dust polarized emission toward star forming regions
trace the magnetic field component in the plane of the sky and provide
constraints to theoretical models of cloud collapse.}
{We compare high-angular resolution observations of the submillimeter
polarized emission of the low-mass protostellar source NGC~1333 IRAS~4A
with the predictions of three different models of collapse of
magnetized molecular cloud cores.}
{We compute the Stokes parameters for the dust emission for the three
models. We then convolve the results with the instrumental response of
the Submillimeter Array observation toward NGC~1333 IRAS~4A. Finally,
we compare the synthetic maps with the data, varying the model
parameters and orientation, and we assess the quality of the fit by a
$\chi^2$ analysis.}
{High-angular resolution observations of polarized dust emission can
constraint the physical properties of protostars. In the case of
NCC~1333 IRAS~4A, the best agreements with the data is obtained for
models of collapse of clouds with mass-to-flux ratio $>2$ times 
the critical value, initial
uniform magnetic field of strength $\sim 0.5$~mG, and age of the order
of a few $10^4$~yr since the onset of collapse.  Magnetic dissipation,
if present, is found to occur below the resolution level of the
observations. Including a previously measured temperature profile
of \src\ leads to a more realistic morphology and intensity
distribution. We also show that ALMA has the capability of
distinguishing among the three different models adopted in this work.}
{Our results are consistent with the standard theoretical scenario for
the formation of low-mass stars, where clouds 
initially
threaded by large-scale magnetic fields
become unstable and collapse,
trapping the field in the nascent protostar and the
surrounding circumstellar disk. In the collapsing cloud, the dynamics is
dominated by gravitational and magnetic forces.}


\maketitle

\section{Introduction\label{intro}}

Magnetic fields play an important role in the star formation process.
Molecular clouds are expected to form dense cores through a combination
of loss of magnetic and turbulent support. Eventually, a molecular cloud core
overcomes magnetic support (``supercritical'' stage), and collapses
gravitationally. The magnetic field is then pinched and strengthened in
the central regions of the core, and is expected to assume an hourglass
shape (Fiedler et al.~1993; \citealp{galli93a,galli93b,nakamura05}).

Aspherical spinning dust particles tend to align their small axis
parallel to the direction of the magnetic field. Thermal emission from
such elongated grains is thus partially linearly polarized, with the
polarization vector perpendicular to the magnetic field. Consequently,
the polarized emission is a good tracer of the magnetic field.  To test
the influence of magnetic fields we compare high-angular resolution
observations of the polarized emission measured at submillimeter
wavelengths toward the low-mass protostar NGC~1333~IRAS~4A with 
non-turbulent magnetohydrodynamic (MHD) models of molecular cloud cores
threaded by an initial uniform magnetic field.  This first step will
help to ({\em i}\/) select the best models to describe the structure
and evolution of low-mass cores, and, ({\em ii}\/) to better understand
the importance of the physical processes involved in their formation
and evolution. In a subsequent paper we will consider models of
magnetized molecular cores  formed in a turbulent environment.

The low-mass protostar \src\ is an ideal test site for models of magnetized
cloud collapse and star formation.  BIMA spectropolarimetric observations at
1.3~mm have detected and partially resolved the polarization in both the dust
and CO~(2--1) emission \citep{girart99}, showing hints of a hourglass morphology
of the magnetic field. Recent polarimetric observations with the SMA at
877~$\mu$m with a resolution of 1\asec3 (390~AU) have shown that the magnetic
field associated with the infalling envelope has a clearly ``pinched''
morphology on a scale of a few hundreds AU (see Fig.~1 in \citealt{girart06}).
This morphology resembles the hourglass
shape that is predicted by the standard theory of low-mass star formation in a
collapsing core with a regular magnetic field dominating the irregular
(turbulent) one \citep{fiedler93,galli93a,galli93b,nakamura05}.
Applying the Chandrasekhar-Fermi equation,
\citet{girart99} derived a magnetic field strength in the plane of the sky
(POS) of $B_{\mathrm{POS}}\approx 5$~mG, corresponding to a mass-to-flux ratio
of $\sim 1.7$ times the critical value.

\citet{goncalves08} compared the position angles in the plane of the
sky of the polarization vectors determined by \citet{girart06} with the
inclination of magnetic field lines of ideal \citep{galli93a,galli93b}
and non-ideal \citep{shu06} MHD collapse models.  They found a good
qualitative agreement for a source with $\lesssim 1$~$M_\odot$ and a
mass-to-flux ratio of $\sim 2$ times the critical value. The present
work is a step forward in the modelization and methodology with respect
to that of \citet{goncalves08}.

This paper is organized as follows: in sections~\ref{sec_iras4a} and
\ref{models} we describe the target source \src\ and the selected MHD
models, respectively. In section~\ref{method} we describe the synthetic
map generation. General results are detailed in
section~\ref{sec_results}. In section~\ref{sec_models} we present the
MHD models prediction convolved with the SMA interferometer and compare
them with \src\ observations. In section~\ref{sec_alma} we present the
ALMA maps of the MHD model prediction. Finally, in
section~\ref{summary} we summarize the results and list the
conclusions.

\section{NGC~1333~IRAS~4A}
\label{sec_iras4a}

\begin{figure*}[ht]
\includegraphics[width=18.4cm,angle=0]{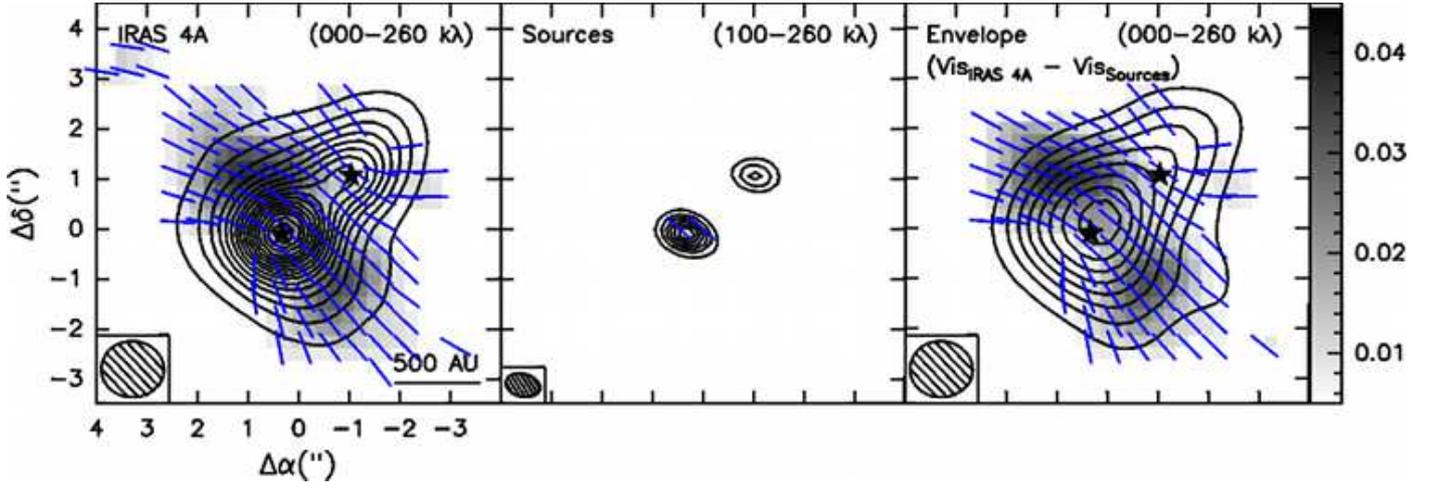}
\caption{{\it Left panel}: \src\ combining SMA sub-compact, compact and
extended configurations with robust parameter set to 0.5. The stars
mark the two compact sources (see central panel). Top-right corner
shows the {\it u,v} range used. The synthesized beam is
1\farcs24 $\times$ 1\farcs12. Contours show the dust emission at
880~$\mu$m in steps of $6\sigma$, from $6\sigma$ to $96\sigma$, where
$\sigma=0.02$~Jy~beam$^{-1}$. The pixel map shows the polarized
intensity (see scale on the right-hand side). Blue bars show the
observed magnetic field direction. {\it Central panel}: \src\ compact
components. Legends are the same as in the left panel. The {\it u,v}
range is restricted to the longest baselines. The synthesized beam is
0\farcs70 $\times$ 0\farcs46. Contours range from $6\sigma$ to
$36\sigma$. {\it Right panel}: \src\ envelope emission (see
section~\ref{sec_iras4a_new}). Legends are the same as in the left
panel.  Contours range from $6\sigma$ to $48\sigma$.}
\label{fig_iras4a} 
\end{figure*}

The Perseus molecular cloud is an active low-mass star forming region,
located at a distance ranging from 230~pc to 350~pc \citep{ridge06}.
For this work, we adopt the value of 300~pc \citep{girart06}. In the
southern part of the reflection nebulae NGC~1333, \citet{jennings87}
were the first to identify the protostar NGC~1333~IRAS~4.
\citet{sandell91} resolved the system into two different components,
IRAS~4A and IRAS~4B, separated by $\sim$31\arcsec. They measured a
luminosity of $\sim 28$~$L_\odot$ (at 350~pc, $11~L_\odot$ at 220~pc)
equally shared between the two components. Subsequent interferometric
observations have revealed further multiplicity: IRAS~4A is itself a
binary system. The two components IRAS~4A1 and IRAS~4A2 are separated
by 540~AU (1\farcs8, \citealt{lay95,looney00,girart06}).

This low-mass stellar system is in a very early stage of evolution.
IRAS~4A and 4B are still embedded in an dense molecular and dusty
envelope. \citet{sandell91} derived from submillimetric continuum
single-dish observations a mass of $\sim 9$~$M_\odot$. Subsequent
interferometric observations derived a mass of 1.2~$M_\odot$
\citep{girart06} for the compact component. \citet{difrancesco01}
detected infall motions from inverse p-Cygni profiles observed in
H$_2$CO~(3$_{12}$--2$_{11}$) and N$_2$H$^+$~(1--0).  Single-dish CO
(3--2) observations revealed a NE-SW well-collimated outflow arising
from IRAS~4A \citep{blake95}. \citet{choi05} reports, through
interferometric SiO~(1--0) observations, a highly collimated NE-SW
outflow with a projected position angle of $\sim 19^\circ$, and hints
of a N-S outflow. The author proposes that IRAS~4A2 is powering the
main outflow while IRAS~4A1 would power the secondary one.

\subsection{New data}
\label{sec_iras4a_new}

For this work we generated new observational maps of \src\ combining
compact \citep{girart06}, and sub-compact and extended (Ching \& Lai,
priv.\ comm.) configuration SMA data (see left panel of
Fig.~\ref{fig_iras4a}), consisting of 8~h tracks in polarization mode
at 880~$\mu$m. The data reduction was performed using MIRIAD, while the
imaging was done using GREG from the GILDAS package. To obtain the maps
we used a weighting robust parameter of 0.5 \citep{briggs95}
corresponding to a beam of 1\farcs24 $\times$ 1\farcs12, slightly smaller than
that of \citet{girart06}.  Adding the sub-compact configuration
improved the map with respect to that of \citet{girart06}: the sampling
of the larger scales is better and allows a better characterization of
the circumbinary envelope. In addition, the extended configuration data
help in separating the emission arising from either the embedded
compact sources or the circumbinary envelope. The combined continuum
(Stokes {\it I}) map is in good agreement with that of
\citet{girart06}, atlhough the emission is more extended and has a
sharper morphology.  Furthermore, the polarized intensity map covers a
larger area, has a slightly higher intensity peak and a more defined
morphology.

The dust emission of IRAS~4A arises from the cold circumbinary envelope
and from the warm circumstellar material around each protostar. Since
the focus of this paper is on the morphology of the magnetic field in
the circumbinary envelope, we have subtracted the contribution from the
circumstellar component to the SMA visibility data.  To do so, we first
derived a map of the longest baselines (100-260~$k\lambda$)
corresponding to a beam of 0\farcs70 $\times$ 0\farcs46. At these {\it
u,v}-distances, the emission from the circumbinary envelope is resolved
out, and the only contribution from the dust emissions arises from the
circumstellar material (see central panel of Fig.~\ref{fig_iras4a} and
Fig.\ref{fig_uv-I}). Then, the clean components of this map were
subtracted from the original visibilities. Finally, we obtained a new
map of the circumbinary envelope using the resulting visibilities (see
right-hand side panel of Fig.~\ref{fig_iras4a}).

Table~\ref{tab_param_obs} shows the main parameters of the emission associated
only with  the circumbinary envelope: peak position, $T_{\rm dust}$, RMS,
$S_\nu$, $I_\nu^{\rm{peak}}$, FWHM, $N_{\rm H_2}$, $n_{\rm H_2}$ and mass. The
integrated flux is 4.1~Jy, corresponding to a mass of 0.8~$M_\odot$, both
slightly smaller than those of \citet{girart06} as we were able to isolate the
envelope.  The optical depth of the dust emission at 880~$\mu$m imply that the
observations trace very deep into the source. 
Therefore, neglecting 
scattering (see Sect.~\ref{assumptions}) and assuming an anisotropic radiation
field, the polarized dust
emission is probably originated in the alignment of
dust grains to the magnetic field (see \citealp{lazarian03}, for a 
comprehensive review on this topic).
The envelope hourglass morphology of the
magnetic field is more evident than in earlier data. A new feature is the
double peak in polarized intensity. The map also shows a significant
depolarization toward the source main axis, which was not as clear in the
\citet{girart06} map. This feature can be explained in terms of projections
effects intensified by beam smearing (see, e.g., \citealp{goncalves05}).

\begin{table}[t]
\caption{IRAS~4A: envelope continuum emission at 880~$\mu$m 
and derived parameters~$(^{\rm a})$.}
\begin{tabular}{ll}
\hline\hline
$\alpha({\rm J2000})$~$(^{\rm b})$	& 3h~29m~10.520s \\
$\delta({\rm J2000})$~$(^{\rm b})$	&$31^\circ$~$13^\prime$~$31^{\prime\prime}.12$ \\
$T_{\mathrm{dust}}$~$(^{\rm c})$ & 50~K \\
\hline\smallskip
RMS$_\mathrm{I}$			& 20~mJy~beam$^{-1}$ \\
RMS$_\mathrm{Q}$			& 2.5~mJy~beam$^{-1}$ \\
RMS$_\mathrm{U}$			& 2.5~mJy~beam$^{-1}$ \\
\hline
$S^\mathrm{I}_{\nu}$			& $4.1 \pm 0.4$~Jy\\
$I^\mathrm{I}_{\mathrm{peak}}$		& $1.03 \pm 0.02$~Jy~beam$^{-1}$ \\
FWHM~$(^{\rm d})$			& 1156~AU ($3\asec85$) \\
\hline
$S^\mathrm{pol}_{\nu}$			& $160  \pm 16$~mJy\\
$I^\mathrm{pol}_{\mathrm{peak}}$	& $38.2 \pm 2.5$~mJy~beam$^{-1}$ \\
$\Omega^\mathrm{pol}$~$(^{\rm e})$	& 14 arcsec$^2$	\\
\hline
$\tau$					& 0.07 \\
$N_{\rm H_2}$~$(^{\rm f})$		& $1.2 \times 10^{24}$~cm$^{-2}$ \\
$n_{\rm H_2}$~$(^{\rm f})$		& $1.1 \times 10^8$~cm$^{-3}$ \\
Mass~$(^{\rm f})$			& 0.8~M$_\odot$ \\
\hline
\end{tabular}
\\
($^{\rm a}$) {See Appendix A of \citet{frau10} for details.} \\
($^{\rm b}$) {From a 2D Gaussian fit to the source.}\\
($^{\rm c}$) {\citet{girart06}.} \\
($^{\rm d}$) {Diameter of the circle with the same area as the region of the source
 with intensity above half of the peak.} \\
($^{\rm e}$) Solid angle of the region 
 with polarized intensity above 3-$\sigma$. \\
($^{\rm f}$) {Assuming $\kappa_{250~{\rm GHz}}=1.5$~cm$^2$~g$^{-1}$ and a gas-to-dust ratio of 100 \citep{girart06}.} \\
\label{tab_param_obs}
\end{table}

\section{Theoretical models}
\label{models}

We compare the dust polarization map of IRAS~4A described in the previous
section with the predictions of three models of magnetized cloud collapse. The
models of \citet{galli93a,galli93b} and \citet{allen03a,allen03b} give the
density profile and magnetic field distribution of an infalling envelope
surrounding a low-mass star, the two models differing mainly in the choice of
the initial conditions. The \citet{shu06} model is similar to the previous two,
but contains a parameter representing the spatial scale where the diffusive
effects associated to an electric resistivity (assumed uniform) dominate the
evolution of the magnetic field. Therefore, our analysis
is not able to test the theory of core formation from iniatially subcritical
conditions by ambipolar diffusion. This can only be accomplished by spatially
resolved Zeeman observations of molecular cloud cores and their surroundings
(see e.g.  \citealp{crutcher09}). In a following paper (Frau et al., in preparation) we will
analyze synthetic polarization maps of protostellar cores extracted from
numerical simulations of turbulent clouds.

\subsection{\citet{galli93a,galli93b}}
\label{model_galli}
				
This model follows the collapse of a singular isothermal sphere
threaded by an initially uniform magnetic field. The cloud is assumed
to be non rotating.  This initial condition is a highly idealized
representation of a non-equilibrium state. Inside the collapse region,
bounded by an outward propagating slow magnetosonic wave, the magnetic
field dragged by the flow (even in the presence of ambipolar diffusion)
deflects the infalling gas towards the midplane, forming a large
pseudodisk. The initial state depends on two dimensional quantities,
$r_0 =2a^2/\sqrt{G}B_0$ and $t_0=2a/\sqrt{G}B_0$, defining the
characteristic spatial and  temporal scale of the collapse.  These
depend on the sound speed $a$,  the gravitational constant $G$ and the
initial (uniform) magnetic strength $B_0$. For given $r_0$ and $t_0$,
the time evolution depends on the  non-dimensional parameter
$\tau=t/t_0$, where $t$ is the time elapsed since the onset of
collapse. Fixing $a=0.35$~km~s$^{-1}$, the model thus depends only on
$B_0$ and $\tau$.

\subsection{\citet{allen03a,allen03b}}
\label{model_allen}

This model is similar to that of \citet{galli93a,galli93b} with some
important differences: ({\it i}\/) being fully numerical, it overcomes
the spatial and temporal limitations of the semi-analytical approach of
\citet{galli93a,galli93b}; ({\it ii}\/) the initial state is a
magnetostatic unstable equilibrium configuration (a ``singular
isothermal toroid'' see \citealp{li96}), already flattened in the
direction perpendicular to a magnetic field possessing a hourglass
morphology from the start; ({\it iii}\/) the cloud can rotate around an
axis parallel to the axis of the magnetic field. As in the
\citet{galli93a,galli93b}, magnetic field lines internal to a
``separatrix'' are dragged into the accreting protostar.

The initial configuration is specified by the sound speed, $a$ (as in
the \citealp{galli93a,galli93b} model), and the level of magnetic to
thermal support, $H_0$, which represents the fractional overdensity
supported by the magnetic field above that supported by the thermal
pressure, and the rotational speed, $v_0$. The parameter $H_0$ is
related to the mass-to-flux ratio of the cloud.

The flattening of the mass distribution (the ``pseudodisk'') and the
magnetic field geometry are little affected by rotation. Conversely,
the angular velocity of the infalling gas is strongly influenced by the
magnetic braking associated to the strong field created by accretion,
assuming ideal MHD. This effect has important implications for the
formation of rotationally supported disks around young stars (see
\citealp{galli06}).

\subsection{\citet{shu06}}
\label{model_shu}

To overcome the difficulties associated to catastrophic magnetic braking 
and to the huge magnetic flux of the protostar, \citet{shu06} consider the
consequences of non-ideal MHD effects during the accretion phase of
low-mass star formation. In steady state, magnetic dissipation occurs
inside a region of radius equal to the so-called ``Ohm radius'',
$r_{\rm{Ohm}}=\eta^2/(2GM_\star)$, where $\eta$ is the Ohmic
resistivity (assumed uniform), $G$ is the gravitational constant, and
$M_\star$ is the mass of the accreting protostar.  Outside $r_{\rm
Ohm}$, the accreting gas is in free fall along radial field lines, that
become straight and uniform inside $r_{\rm Ohm}$. The magnetic flux
accreted by the central protostar is zero at all times.

\section{Synthetic map generation}
\label{method}

\subsection{Assumptions}
\label{assumptions}

To compare the intensity and the polarized intensity predicted by the
models with the observed data, it is important to consider the effects
of a temperature gradient, since the sub-mm emission is roughly
proportional to the temperature.  In this work, we have assumed both a
uniform temperature profile (UTP) and a radial temperature profile
(RTP) derived for IRAS~4A by \citet{maret02} from water emission.
Although the theoretical models considered here are computed assuming
an isothermal equation of state, the \src\ observed temperature
gradient does not significantly affect the dynamics of collapse,
because the kinetic energy due to thermal motions is more than one order 
of magnitude smaller than the kinetic energy of the infalling particles.
For example, the temperature expected at 600~AU  is 50~K, which
leads to a thermal broadening of $\sigma_{\rm therm}$$\sim$0.4~km~s$^{-1}$,
whereas the infall velocity expected is $v_{\rm ff}$$\sim$1.7~km~s$^{-1}$. 

We consider optically thin emission with no absorption or scattering
effects, in agreement with the sub-mm emission properties
\citep{hildebrand83,novak89} and with the opacity derived in \src\ (see
Table~\ref{tab_param_obs}).  We have assumed uniform grain properties,
represented by the polarizing
efficiency parameter $\alpha$ which
includes the absorption cross section and the alignment efficiency.  
Following \citet{fiege00}, the maximum degree of polarization is
\begin{equation}
p_{\rm max}=\frac{\alpha}{1-\alpha/6}.
\end{equation}
We assumed $p_{\rm max}=15$\%, corresponding to $\alpha=0.15$.
Despite the high value of $\alpha$ used, and the fact that the grain
properties may change with density \citep{fiege00}, we find that the
absolute polarized intensities derived in the models match reasonably
well the observed values in \src. Lower values of $\alpha$ (e.g.
$\alpha=0.1$) did not reproduce the data equally well.

We performed the numerical integration using an equally spaced regular
cubic grid and uniform step in the line-of-sight direction.

\subsection{Method}
\label{synth_maps}

We improved the technique developed by \citet{goncalves08} to compare
theoretical models with observed data, including in the process the
instrumental effects. In practice, we simulated all the steps of a
regular observing run with both SMA and ALMA, generating synthetic maps
with the same filtering and processing as the observed maps. With this
technique we avoid any possible misinterpretation due to the effects of
the instrumental response and filtering, as well as the data
modification because of the Fourier transform of the observed
visibilities and the subsequent application of the dirty map cleaning
algorithm. The process consists of a series of 5 consecutive steps for
each realization:

\begin{enumerate}

\item  For any given model we generated three-dimensional (3D) data
cubes of density and magnetic field components. The orientation in
space of the 3D source models were defined by two viewing angles:  the
position angle $\phi$ of the projection of the polar axis in the plane
of the sky with respect to the north direction, and the inclination
angle $\omega$ of the polar axis with respect to the plane of the sky
($\omega =0^\circ$ for edge-on view).  Since the models used have axial
symmetry, the optically thin emission assumption allowed us to explore
only half of the inclination angle space ($0^\circ\leq\omega\leq
90^\circ$).  We restricted $\phi$ to the range $0^\circ\leq\phi\leq
90^\circ$ (the observations fix the magnetic axis of \src\ at
$\phi\approx 50^\circ$, see \citealp{girart06}).

\item In the plane of the sky we simulated a square area with side
length of 51\farcs2 ($\sim 1.5\times 10^4$~AU at the distance of
IRAS~4A). The map size was chosen to be about twice the SMA primary
beam to better process the sidelobes in the final maps. In this plane
we used a grid of $512\times 512$ pixels with a pixel size of
0\farcs1 ($\sim 30$~AU), enough to oversample the smallest beam used
in this work ($\sim0.4$\arcsec $\sim120$~AU for ALMA).  With this
choice we ensured, in the final convolved maps, the independence of
points separated more than a beam distance due to beam convolution.  In
the line-of-sight direction we covered a length of $6\times 10^3$~AU
(equivalent to 20\arcsec) sampled with 60 cells.  A larger integration
length or a larger number of steps did not affect significantly the
details of the final maps.

\item Through a ray-tracing scheme, we integrated the emission of the
cells along the line-of-sight $\ell$ generating 2D raw synthetic maps
for the Stokes parameter $I$, $Q$, and $U$. We followed a method
developed by \citet{lee85}, and elaborated by \citet{wardle90},
\citet{fiege00}, and \citet{padoan01}. One can calculate the Stokes
$Q$ and $U$ intensities as $Q=Cq$ and $U=Cu$, where $C$ is a constant
that includes all the terms assumed to be constant (polarization
efficiency and polarization and absorption cross sections) that can be
interpreted as a polarized intensity scale factor. $q$ and
$u$ are the ``reduced'' Stokes parameters defined as

\begin{equation}
q=\int{\rho B_\lambda\left( T_\mathrm{d}\right)\ \cos{2\psi}\cos^2{\gamma}\ d\ell},
\end{equation}
\begin{equation}
u=\int{\rho B_\lambda\left( T_\mathrm{d}\right)\ \sin{2\psi}\cos^2{\gamma}\ d\ell},
\end{equation}

where $\rho$ is the density, $B_\lambda(T_{\rm d})$ is the Planck
function at the dust temperature $T_{\rm d}$, $\psi$ is the angle
between the north direction in the plane of the sky and the component of
$\vec{B}$ in that plane, and $\gamma$ is the angle between the local
magnetic field and the plane of the sky.  
Stokes $I$ is given by $I=(C/\alpha)(\Sigma-\alpha\Sigma_2)$
\citep{fiege00} where $\alpha$ is the maximum polarizing efficiency
assumed to be 15\%, the $C/\alpha$ factor can be interpreted as a
total intensity scale factor, while $\Sigma$ and $\Sigma_2$ are defined
as

\begin{equation}
\Sigma=\int{\rho B_\lambda\left( T_\mathrm{d}\right)\ d\ell},
\end{equation}
\begin{equation}
\Sigma_2=\int{\rho B_\lambda\left( T_\mathrm{d}\right)\ 
\left(
\frac{\cos^2{\gamma}}{2}-\frac{1}{3}
\right)
\ d\ell},
\end{equation}

and represent the emitted Stokes $I$ intensity and the
polarization absorption losses respectively. As for Stokes $Q$ and $U$,
one can define the "reduced" Stokes $I$ parameter as

\begin{equation}
i=\frac{\Sigma-\alpha\Sigma_2}{\alpha}.
\end{equation}

\item Once the synthetic Stokes 2D maps were generated, the flux was
rescaled so that the total flux in the region of the map with
$I_\nu>6\sigma$ (see section~\ref{sec_iras4a_new}) matched that of the
synthetic map. A Gaussian noise was added to match the noise of the
observations for each Stokes map. These maps were converted to
visibilities using the same visibility coverage in the $u,v$ plane as
the real observations. In this step we mimicked the effects of the
observation noise and the instrumental filtering. This allowed us to be
sensitive to the same spatial scales and to similar emission levels. 
Then, the final model maps for the Stokes $I$, $Q$ and $U$ were
obtained from the synthetic visibilities in the same way as the
combined SMA maps presented in Fig.~\ref{fig_iras4a}.

\item The process described in points 1 to 4 is repeated a large number
of times for both temperature treatments using different values for
({\it i}\/) the position angle $\phi$ and the inclination angle
$\omega$, and ({\it ii}\/) the model parameters.

\end{enumerate}

For ALMA, we used the task {\it simdata} from the CASA package to predict the
expected maps for the models with the ALMA capabilities.  At the source
distance, the cells of the simulations had a typical length of $\sim
30$~AU. We chose the full ALMA configuration~09, which provides a
synthesized beam of 0\farcs7 $\times$ 0\farcs4 ($210 \times 120$~AU)
thus ensuring that the final synthetic maps were not affected by
resolution issues. We simulated a 2~hr run at 345.8~GHz in polarization
mode. As good weather is required for polarization measurements, we
assumed 1~mm of precipitable water vapor. The elevation of the source
ranged between 30$^\circ$ and 40$^\circ$.

\subsection{Selection}

At this point, one has to select the ``best'' procedure for comparing
the synthetic maps with the observational data.  This can be
accomplished in several ways: (1) using the method of
\citet{goncalves08} based on the minimization of the difference between
the observed and predicted position angles of the polarization vectors
(hereafter simply ``angle difference method''); (2) performing a
$\chi^2$ analysis of the synthetic Stokes $Q$ and $U$ maps with respect
to the observed maps. In the latter case (hereafter simply ``$\chi^2$
method''), we positioned the peak of the synthetic map on the peak of
the \src\ envelope (see Fig.~\ref{fig_iras4a}), and we compared the
synthetic map with all the region of the observed map with intensity
larger than $3\sigma$ (see Table~\ref{tab_param_obs}). As we focus on
the polarized emission, we define the best fitting models as those
which minimized the sum $\chi^2=\chi_Q^2+\chi_U^2$. Stokes $I$ was
excluded since it shows considerable dependence on the assumed
temperature profile.

\begin{figure}[ht]
\centering
\includegraphics[width=9.5cm,angle=0]{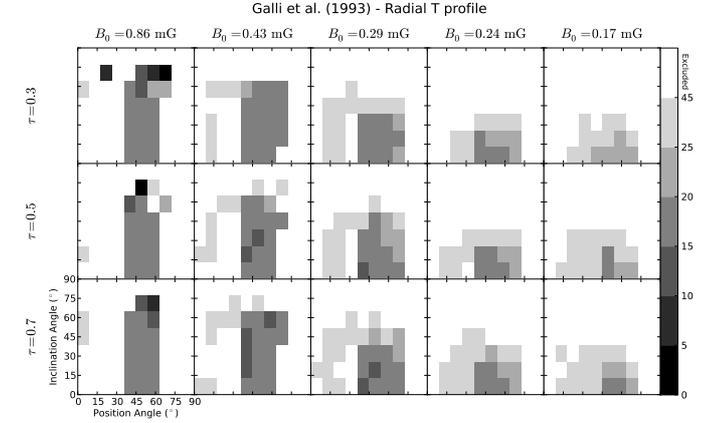}
\caption{{\it Pixel map}: Standard deviation of the difference of the
synthetic polarization vectors with respect to the observed ones for the
\citet{galli93a,galli93b} model with radial temperature profile. White
pixels represent excluded models (see text). Grayscale is on the
right-hand side of the panel.
}
\label{fig_galli93T_pa_sigma_fil}
\end{figure}

\begin{figure}[ht]
\centering
\includegraphics[width=10cm,angle=0]{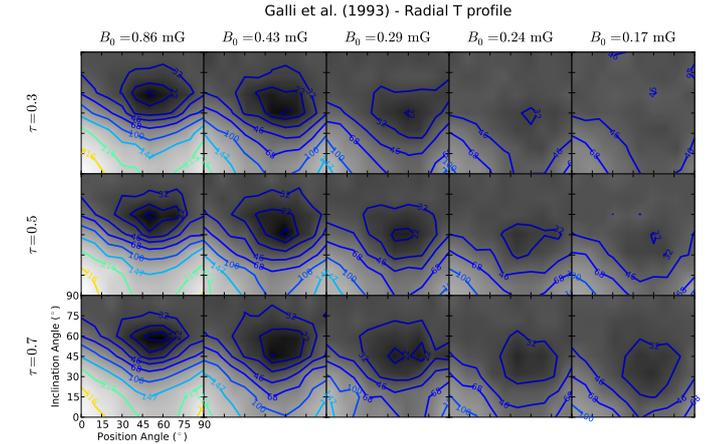}
\caption{{\it Pixel map}: Sum of the $\chi^2$ of the Stokes $Q$ and $U$
maps for the \citet{galli93a,galli93b} model with radial temperature
profile with respect to  the data.  {\it Contours}: 10, $10^{1/6}$,
$10^{2/6}$, \ldots, $10^3$ $\chi^2$ levels.}
\label{fig_galli93T-chi}
\end{figure}

We illustrate the results for the two selection methods for the
\citet{galli93a,galli93b} models with RTP (see
Figs.~\ref{fig_galli93T_pa_sigma_fil} and \ref{fig_galli93T-chi}).
Fig.~\ref{fig_galli93T_pa_sigma_fil} shows the standard deviation of
the distribution of differences in  position angles as function of the
viewing angles $\omega$ and $\phi$ for the Galli \& Shu~(1993a,b)
models described before. Only runs with average difference value of the
position angles with respect to those of \src\ lower than 15\deg\ are
shown. Best-fitting models are characterized by the smallest values of
the standard  deviation. The uniform distribution of results makes
evident the low discrimination power of the angle difference method.
Conversely, the $\chi^2$ method allows to perform a more significant
selection of the best-fitting models.  Fig.~\ref{fig_galli93T-chi}
shows a difference of more than one order of magnitude between bad- and
well-fitting runs thus providing a higher discriminating power among
all the runs. Therefore, in the rest of this paper, we adopt the
$\chi^2$ selection method.

\section{Results}
\label{sec_results}

\subsection{Orientation angles}
\label{res_rot}

\begin{landscape}
\begin{figure}
\centering
\includegraphics[width=24.5cm,angle=0]{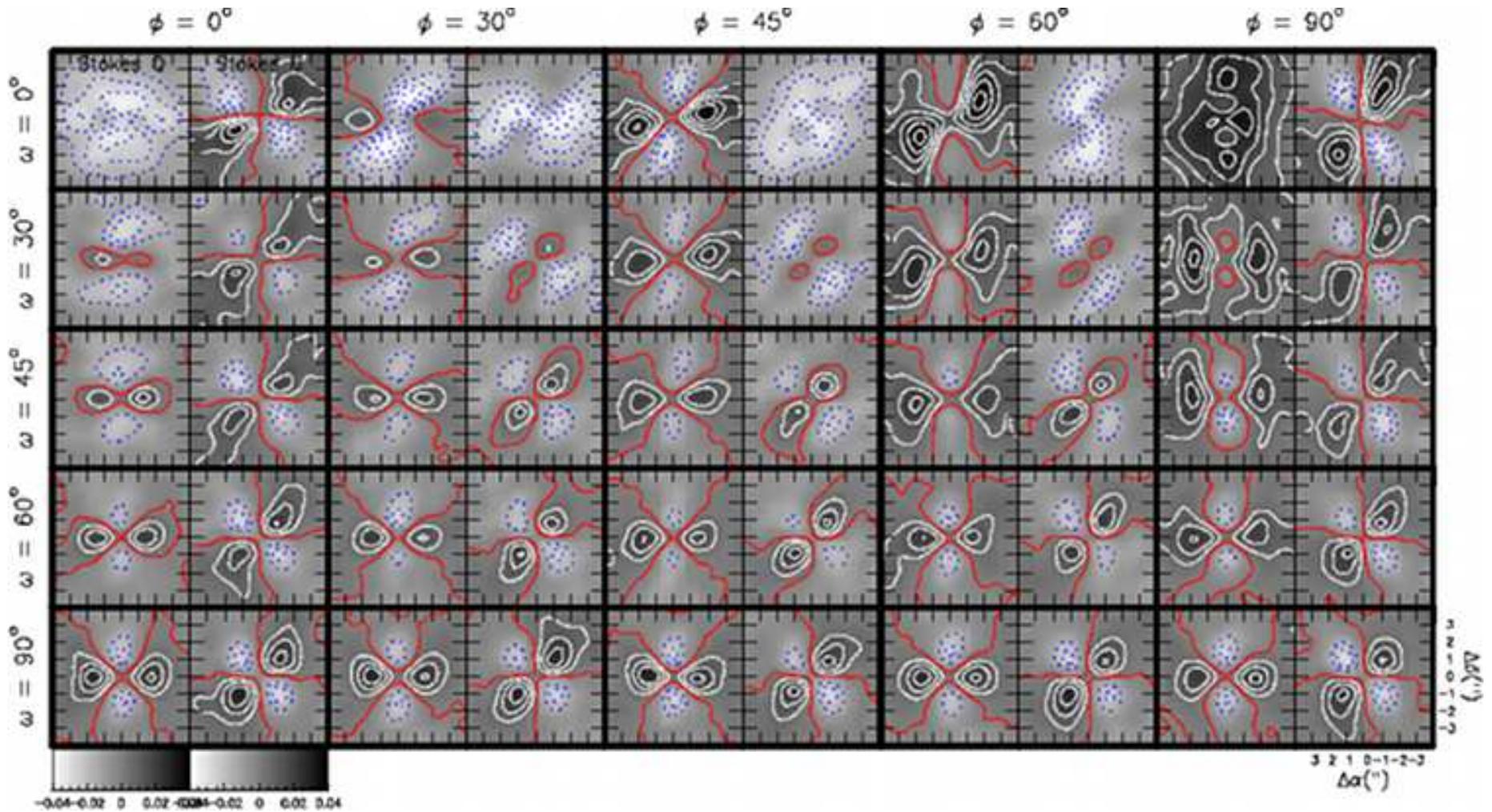}

\caption{ Dependence of the synthetic Stokes $Q$ and $U$
maps on the orientation angles. The maps shown corresponds to the
\citet{allen03a,allen03b} model with $H_0=0.125$, $v_0=0$ and
$t=2\times10^4$~yr. {\it Thick contours}: correspond to a single
combination of position and inclination angle. {\it Individual panels}:
Inside thick contours two panels are shown.
Stokes $Q$ and $U$ map are
shown in the left- and right-hand side panels, respectively. Common
color scale for each Stokes map is shown below the first column. The
angular scale is shown in the bottom right-hand side panel.  {\it Map
contours}: 
represent steps of 3-$\sigma$ starting at 3-$\sigma$, where
$\sigma$=2.5~mJy~beam$^{-1}$. {\it Columns}:  correspond to a position
angle, $\phi$, shown on the top. {\it Rows}: correspond to an
inclination angle, $\omega$, shown on the left-hand side of the
figure.}

\label{fig_rot_QU}
\end{figure}
\end{landscape}

\begin{figure*}[ht]
\centering
\includegraphics[width=\textwidth,angle=0]{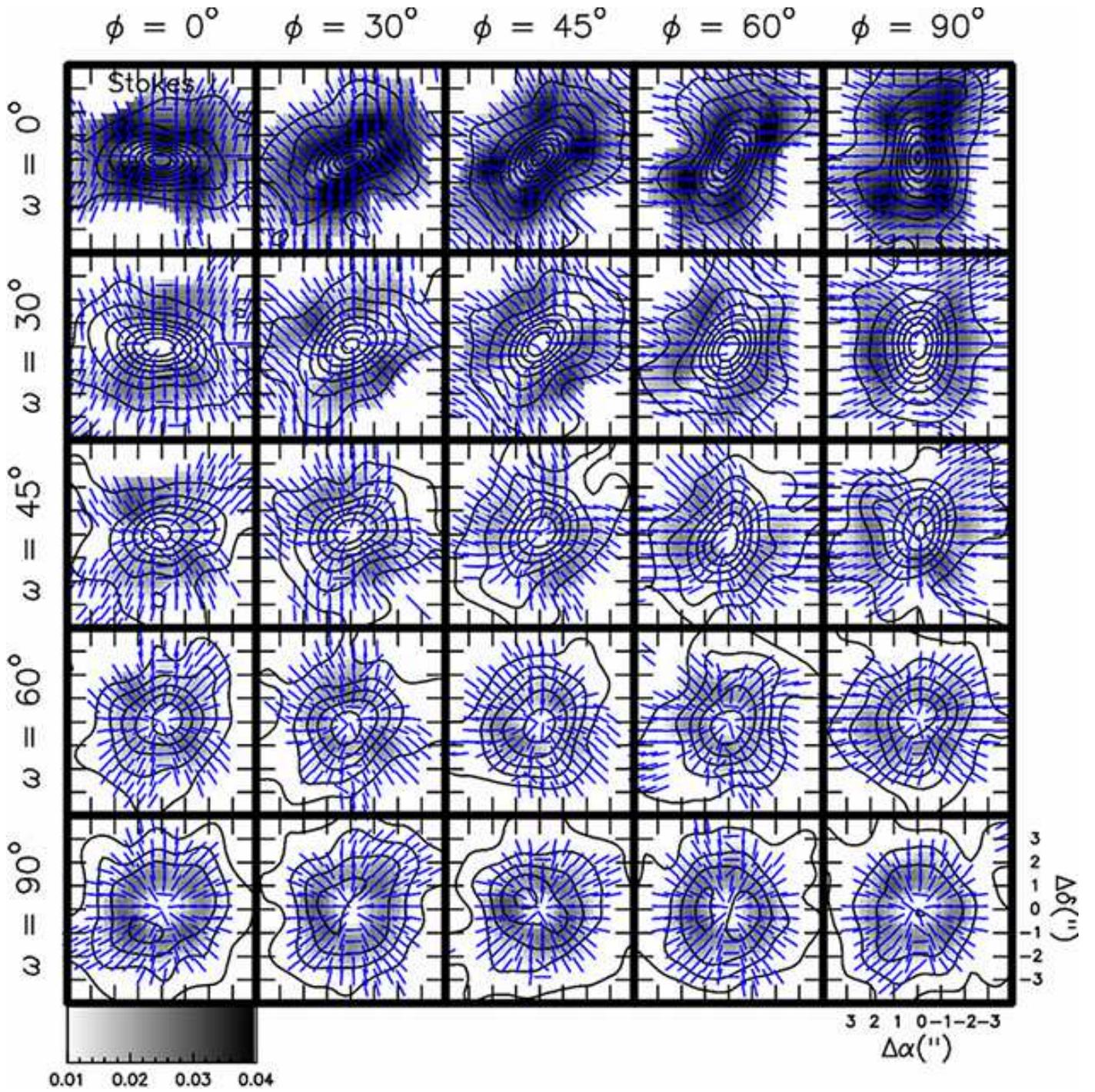}

\caption{
Same as Fig.~\ref{fig_rot_QU} but for Stokes~$I$ maps. 
Each panel show the total emission (contours), the polarized emission (pixel map)
and the magnetic field direction (segments).
{\it Map
contours}: for Stokes~$I$ represent steps of 6-$\sigma$ starting at
6-$\sigma$, where $\sigma$=0.02~Jy~beam$^{-1}$.
}
\label{fig_rot_I}
\end{figure*}

Stokes $Q$ and $U$ maps show a significant dependence on the
orientation angles of the source, and can be used to constrain the
viewing geometry. Fig.~\ref{fig_rot_QU} illustrates the different emission
patterns arising from an \citet{allen03a,allen03b} source, with
$H_0=0.125$, $v_0=0$, and $t=2\times10^4$~yr, after varying the
orientation angles. Stokes $Q$ and $U$ maps are shown for all the
combinations of position ($\phi$) and inclination ($\omega$) angles of
0\deg, 30\deg, 45\deg, 60\deg,  and 90\deg.  Conversely, 
Fig.~\ref{fig_rot_I} shows Stokes $I$
maps which depend marginally on the position angle, and only for small
inclination angles.

\clearpage

\clearpage

As can be seen in Fig.~\ref{fig_rot_QU}, Stokes $Q$ is specially sensitive
to the position angle for small inclination angles, while Stokes $U$
appear to vary more with inclination angle at non-extreme position
angles. From a practical point of view, one could identify with
relatively high precision, by pure comparison with Figs.~\ref{fig_rot_QU}
and \ref{fig_rot_I},
both position angles of a source with a magnetic field with hourglass
morphology and an inclination angle smaller than $\sim 60$\deg.  For
inclination angles $\gtrsim 60$\deg the expected magnetic field tend to
be mostly radial and the Stokes $Q$ and $U$ maps show very similar
morphologies independently of $\omega$. Therefore, polarization
observations have shown to be a powerful tool to determine both
position angles of a sources with respect to the LOS direction.
Figures~\ref{fig_rot_QU} and \ref{fig_rot_I} could be used as 
templates for future
observations of the dust polarized emission toward star forming cores.

\subsection{Temperature profiles}
\label{res_temp}

Figure~\ref{fig_uv-I} shows the differences in Stokes~$I$ {\it u,v}
amplitude between both temperature treatments. The models show more
realistic amplitudes using the RTP. The models with UTP show lower
intensities than \src\ in the $\gtrsim 25$~$k\lambda$ range
($\lesssim 4$\arcsec), while the models with RTP fit the observations up to
15-20~$k\lambda$  ($\lesssim 5$\asec5). Note that this is remarkable for
either temperature treatment as the intensity rescaling was done in the
$I_{\nu}>6\sigma$ which covered radii $\lesssim$3\arcsec
($\gtrsim 35$~$k\lambda$). Figs.~\ref{fig_best-noT} and \ref{fig_best-T}
show the same information in the image domain for UTP and observed RTP,
respectively, for cases with realistic model parameters and orientation
angles. UTP maps show more extended emission (larger {\it u,v} amplitudes
at short baselines) than \src\ and a lower emission peak evident from the
residual maps. On the other hand, RTP maps show a slightly more extended
morphology than \src\ and realistic intensity peak values. Note that RTP
residual maps tend to show zero emission at short radii and slightly
negative emission at large ones due to the more extended sources
predicted by the models. More realistic fluxes, intensity peaks and
masses are derived from the RTP treatment. However, even using the
observed temperatures, the radial intensity profile from models is
steeper than that of \src, which lead to smaller FWHM for model synthetic
maps and, consequently, higher densities than those observed.

The RTP treatment also predicted more realistic Stokes~$Q$ and $U$
maps, shown in the middle left-hand side and middle right-hand side
panels of Figs.~\ref{fig_best-noT} and \ref{fig_best-T}. Although peak
values are similar for both treatments, RTP map show roughly the same
polarized flux over the same solid angle with similar morphology to \src.
On the other hand, UTP map showed roughly twice as much polarized flux
over twice the solid angle of \src\ (see Tables~\ref{tab_models-noT} and
\ref{tab_models-T}). An immediate consequence was the unrealistic UTP
vector map whereas the RTP one reasonably match that of \src.

Summarizing, RTP maps reproduced with higher fidelity the observed
\src\ emission, in the three Stokes parameters, better than UTP maps in
the same conditions. Consequently, the physical parameters derived from
RTP maps were more realistic.

\subsection{Visibility amplitudes}
\label{res_uv}

\begin{figure}[ht]
\centering
\includegraphics[width=9cm,angle=0]{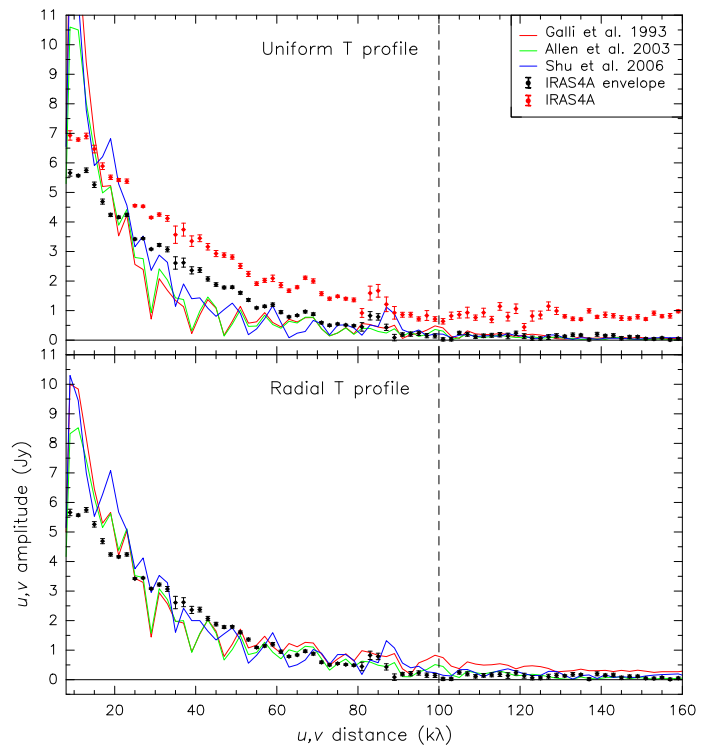}
\caption{Stokes I visibility amplitudes vs. {\it u,v} distance
averaged in bins of 2$k\lambda$.  {\it Top panel}: Uniform temperature
profile. {\it Bottom panel}: Radial temperature profile. {\it Red and
black dots}: \src\ full data and envelope data, respectively, with their
statistical error bars. {\it Vertical dashed line}: {\it u,v} distance
threshold used to derive the compact components map (see
Fig.~\ref{fig_iras4a}). Starting at $\sim$90$k\lambda$ the emission seems
to match with an unresolved source of $\sim$1~Jy. {\it Red, green, and
blue solid curves}: visibility amplitudes derived after convolving the
\citet{galli93a,galli93b}, \citet{allen03a,allen03b}, and \citet{shu06}
models, respectively, used to generate the maps of left-hand panels of
Figs.~\ref{fig_best-noT} and \ref{fig_best-T}. }
\label{fig_uv-I}
\end{figure}

\begin{figure*}[h]
\centering
\includegraphics[height=12cm,angle=0]{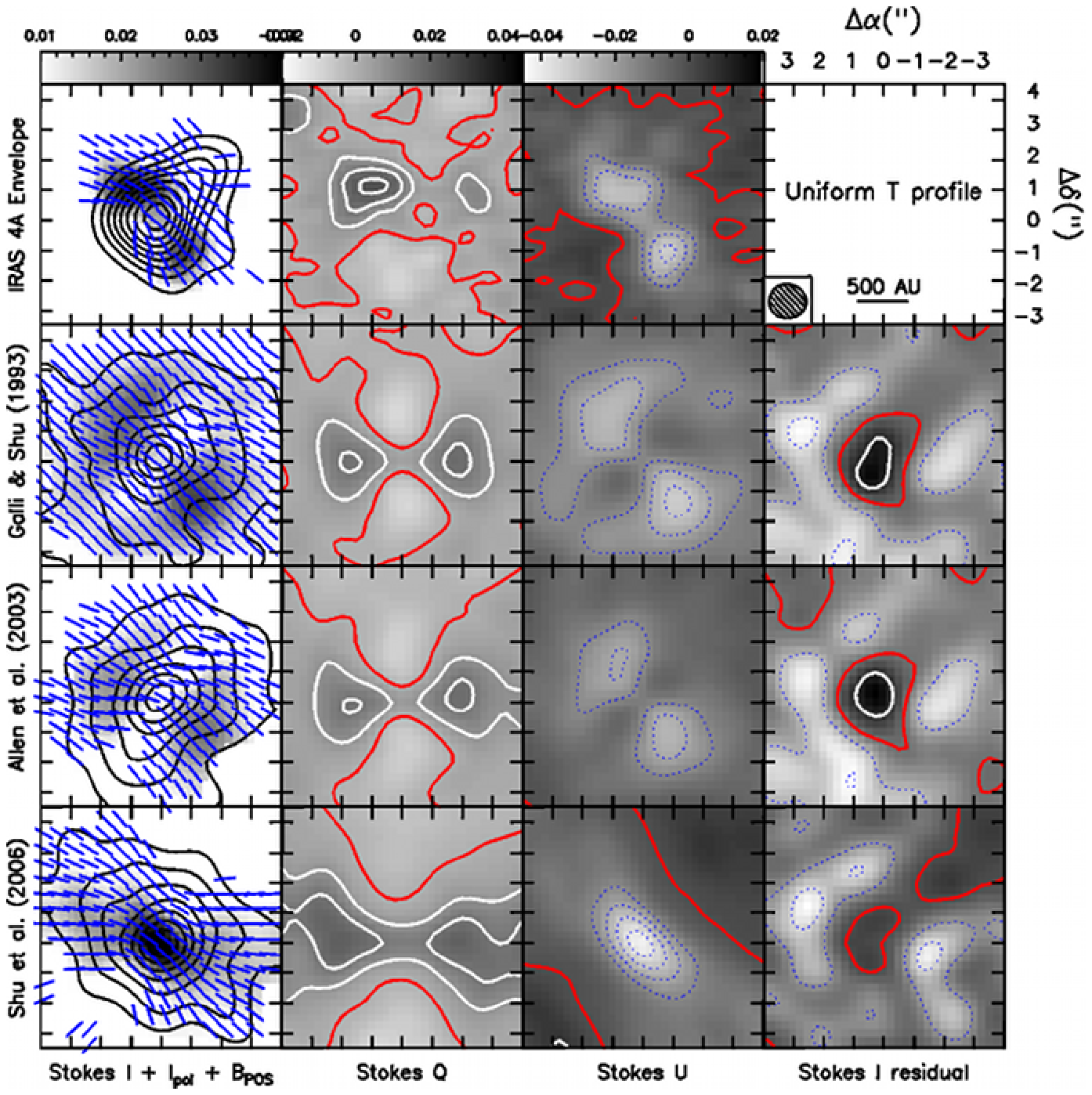}

\caption{Comparison of selected models with data assuming a uniform
temperature of the gas. The orientation angles are fixed to
$\phi$=50$^\circ$ and $\omega$=45$^\circ$ for a better model
comparison. {\it Rows}: IRAS~4A (first row); model
\citet{galli93a,galli93b} with $B_0=0.43$  and $\tau=0.7$ (second row);
model \citet{allen03a,allen03b} with $H_0=0.125$, $v_0=0$ and
$t=10^4$~yr  (third row); and model \citet{shu06} with $r_{\rm
Ohm}=75$~AU (fourth row).  {\it Columns}: In each row, the panels
show:  intensity (first panel, contours), polarized intensity (first
panel, pixel map)  and magnetic field vectors (first panel, segments);
map of Stokes $Q$  (second panel, pixel map and contours); map of
Stokes $U$ (third panel, pixel and contours); residuals models--data
for Stokes $I$ (fourth panel, pixel map and contours). The color scale
is shown on the top of each column. {\it Contours}: Contours for the
Stokes~$I$ maps (left panels) depict emission levels from 6~$\sigma$ up
to the maximum value in steps of 6~$\sigma$, where
$\sigma=0.02$~Jy~beam$^{-1}$. Coutours for the Stokes~$Q$ and $U$ maps
depict levels from the minimum up to the maximum in steps of 3~$\sigma$
where $\sigma=2.5$~mJy~beam$^{-1}$. The solid red contour marks the
zero emission level, solid white contours mark positive emission and
blue dotted contours mark negative emission. Contours for the residual
Stokes~$I$ follow the same rule of Stokes~$Q$ and $U$ but with steps of
6~$\sigma$, where $\sigma=0.02$~Jy~beam$^{-1}$. The top right panel
shows the beam and the angular and spatial scale.}

\label{fig_best-noT}
\end{figure*}

\begin{figure*}[h]
\centering
\includegraphics[height=12cm,angle=0]{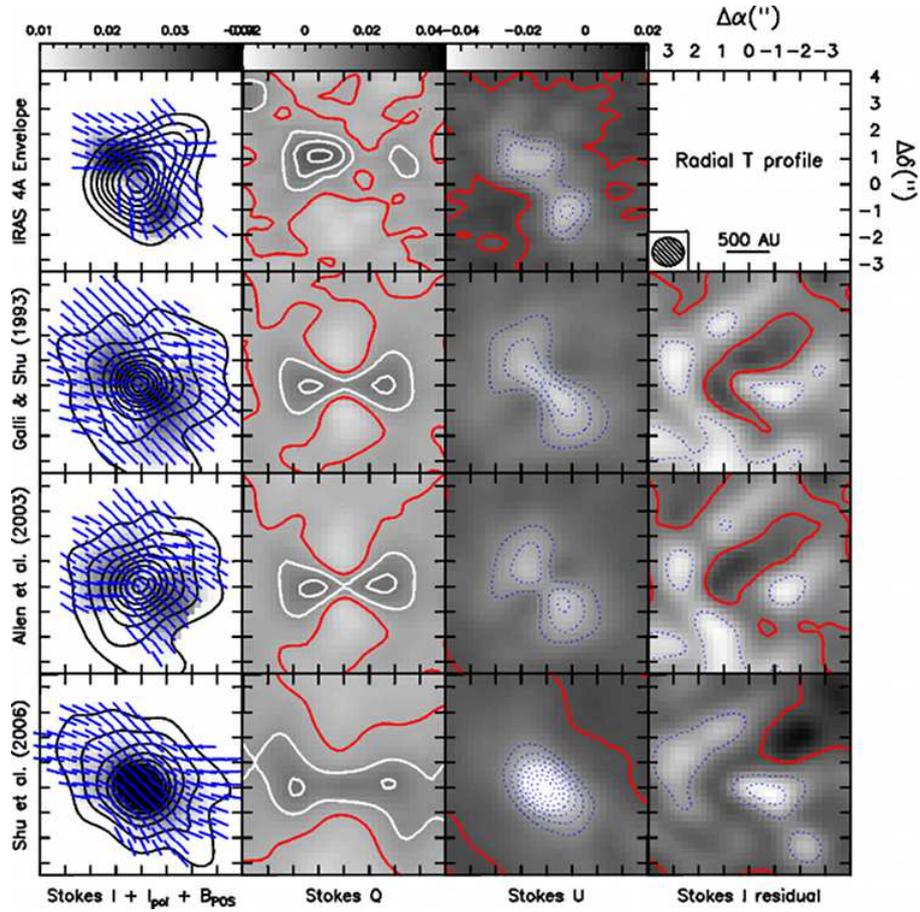}
\caption{ Same as Fig.~\ref{fig_best-noT} for the case with
the \citet{maret02} observational temperature gradient.}
\label{fig_best-T}
\end{figure*}

Solid curves in Fig.~\ref{fig_uv-I} show the total {\it u,v} amplitude
as a function of the {\it u,v} distance for the models using realistic
model parameters and orientation angles ($\phi$=50$^\circ$ and
$\omega$=45$^\circ$, see Sections~\ref{res_rot} and \ref{sec_models}). 
The oscillations in the synthetic visibilities are due to the added
noise. The intensity rescaling of synthetic data was performed in the
image domain as it is the output of the simulations. However, a good
agreement in the {\it u,v} data is important given that it is the
output from the telescope.  Red and black dots in Fig.~\ref{fig_uv-I}
show, respectively, the resulting {\it u,v} amplitudes before and after
subtracting the two circumstellar compact components. We assumed that
only emission from the envelope remained after subtraction.  To test
the goodness of the rescaling method we compared the observed {\it u,v}
data of \src\ with the synthetic {\it u,v} data derived from the
models.  A remarkable agreement (specially using the observed
temperature profile, see Sect.~\ref{res_temp}) is achieved for {\it
u,v} distances ranging from $\sim 20$~$k\lambda$ up to the maximum
baseline with significant envelope emission ($\sim 90$~$k\lambda$).
None of the envelope models showed significant emission at {\it u,v}
distances $\gtrsim 100$~$k\lambda$ (equivalent to a radius of 1" or
300~AU) reinforcing the hypothesis that no envelope emission is
detected from the \src\ envelope in this {\it u,v} range. At {\it u,v}
distances shorter than $\sim$20$k\lambda$ the models show larger
emission than the \src\ envelope. At a distance of 300~pc this scale is
equivalent to emission with a radius of $\gtrsim 1500$~AU ($\gtrsim
5$\arcsec), which is larger than the radius of \src\ at a 3-$\sigma$
level ($\sim 3$\arcsec, $\sim 900$~AU). This excess has its origin in
the fact that the models predict a more extended source than the
observed one, with typical radii of $\sim$4\arcsec ($\sim 1200$~AU) at
a 3-$\sigma$ level.

\begin{table*}[h]
\caption{
Models with uniform temperature profile:
880~$\mu$m continuum emission and derived parameters~$(^{\rm a})$.}
\begin{tabular}{lllll}
\hline
\hline
				& Unit	
	& \citet{galli93a,galli93b} 
				&\citet{allen03a,allen03b}
							&  \citet{shu06}\\
\hline
$S^\mathrm{I}_{\nu}$			& Jy		
	& 8.92 $\pm$ 0.25	& 7.73 $\pm$ 0.25	& 8.35 $\pm$ 0.25	\\
$I^\mathrm{I}_{\mathrm{Peak}}$	& Jy~beam$^{-1}$		
	& 0.77 $\pm$ 0.03 	& 0.79  $\pm$ 0.03	& 0.98 $\pm$ 0.03	\\
FWHM~($^{\rm b}$)		& AU (\arcsec)			
	& 919 (3.06)		& 1015 (3.39)		& 892 (2.84)		\\
\hline
$S^\mathrm{pol}_{\nu}$		& mJy		
	& 390 $\pm$ 40		& 301 $\pm$ 30		& 328 $\pm$ 30		\\
$I^{\rm pol}_{\mathrm{Peak}}$	& mJy~beam$^{-1}$		
	& 31 $\pm$ 2	 	& 34  $\pm$ 2		& 43 $\pm$ 2		\\
$\Omega^\mathrm{pol}$~($^{\rm c}$)	& arcsec$^2$
	& 28			& 30			& 23	\\
\hline
$\tau$				& -				
	& 0.27			& 0.18			& 0.30			\\
$N_{\rm H_2}$~($^{\rm d}$)	& 10$^{24}$~cm$^{-2}$			
	& 4.61			& 3.14			& 5.11			\\
$n_{\rm H_2}$~($^{\rm d}$)	& 10$^{8}$~cm$^{-3}$			
	& 5.03			& 3.10			& 6.02			\\
Mass~($^{\rm d}$)		& $M_{\odot}$			
	& 1.92 			& 1.60			& 1.82			\\
\hline
\end{tabular}
\\
($^{\rm a}$) See Appendix A of \citet{frau10} for details. \\
($^{\rm b}$) Diameter of the circle with the same area as the region of the source
 above the half peak intensity. \\
($^{\rm c}$) Solid angle of the region 
 with polarized intensity above 3-$\sigma$. \\
($^{\rm d}$) Assuming $\kappa_{250~GHz}$$=$1.5~cm$^2$ g$^{-1}$ and a gas-to-dust ratio of 100 \citep{girart06}. \\
\label{tab_models-noT}
\end{table*}

\begin{table*}[h]
\caption{
Models with radial temperature profile:
880~$\mu$m continuum emission and derived parameters~$(^{\rm a})$.}
\begin{tabular}{lllll}
\hline
\hline
				& Unit	
	& \citet{galli93a,galli93b} 
				&\citet{allen03a,allen03b}
							&  \citet{shu06}\\
\hline
$S^\mathrm{I}_{\nu}$			& Jy		
	& 6.59 $\pm$ 0.25	& 6.37 $\pm$ 0.25	& 6.90 $\pm$ 0.25	\\
$I^\mathrm{I}_{\mathrm{Peak}}	$	& Jy~beam$^{-1}$		
	& 1.08 $\pm$ 0.03 	& 1.02  $\pm$ 0.03	& 1.28 $\pm$ 0.03	\\
FWHM~$(^{\rm b})$		& AU (\arcsec)			
	& 693 (2.31)		& 744 (2.48)		& 693 (2.31)		\\
\hline
$S^\mathrm{pol}_{\nu}$		& mJy		
	& 239 $\pm$ 24		& 173 $\pm$ 17		& 230 $\pm$ 23		\\
$I^{\rm pol}_{\mathrm{Peak}}$	& mJy~beam$^{-1}$		
	& 32 $\pm$ 2	 	& 30  $\pm$ 2		& 68 $\pm$ 2		\\
$\Omega^\mathrm{pol}$~$(^{\rm c})$	& arcsec$^2$
	& 16			& 12			& 12	\\
\hline
$\tau$				& -				
	& 0.37			& 0.30			& 0.39			\\
$N_{\rm H_2}$~$(^{\rm d})$	& 10$^{24}$~cm$^{-2}$			
	& 6.28			& 5.12			& 6.64			\\
$n_{\rm H_2}$~$(^{\rm d})$	& 10$^{8}$~cm$^{-3}$			
	& 9.09			& 6.91			& 9.61			\\
Mass~$(^{\rm d})$		& $M_{\odot}$			
	& 1.47 			& 1.39			& 1.57			\\
\hline
\end{tabular}
\\
(a) See Appendix A of \citet{frau10} for details. \\
(b) Diameter of the circle with the same area as the region of the source
 above the half peak intensity. \\
($^{\rm c}$) Solid angle of the region with polarized intensity above 3-$\sigma$. \\
(d) Assuming $\kappa_{250~GHz}$$=$1.5~cm$^2$ g$^{-1}$ and a gas-to-dust ratio of 100 \citep{girart06}. \\
\label{tab_models-T}
\end{table*}

\clearpage

\section{SMA synthetic maps}
\label{sec_models}

In general, synthetic intensity maps obtained from theoretical models
tend to be less concentrated than the observed sources.  The flux
scaling based on the real data, combined with less compact synthetic
sources, cause the synthetic maps to show more extended emission than
the observed ones, and also a lower flux intensity peak.  In the
following subsections (\ref{ssec_galli93}, \ref{ssec_allen03}, and
\ref{ssec_shu06}) we determine the best fitting parameters for each
individual model (see Figs.~\ref{fig_galli93T-chi},
\ref{fig_allen03T-chi}, and \ref{fig_shu06T-chi}) and perform a direct
comparison of \src\ with all the models (see Figs.~\ref{fig_best-noT}
and \ref{fig_best-T}, and Tables~\ref{tab_models-noT} and
\ref{tab_models-T}).

\subsection{\citet{galli93a,galli93b}\label{ssec_galli93}}

We selected 5 values of the initial magnetic field ($B_0= 0.86$~mG,
$0.43$~mG, $0.29$~mG, $0.24$~mG, and $0.17$~mG, corresponding to
$t_0=10^4$~yr, $2\times10^4$~yr, $3\times10^4$~yr, $4\times10^4$~yr, and
$5\times10^4$~yr) and 3 values of the non-dimensional time $\tau$
($\tau=0.3$, 0.5 and 0.7). The mass-to-flux ratio of the initial
configuration is not spatially uniform as in the models of Allen et
al.~(2003a,b) described in Sect.~5.3. A spherical region centered on the
origin and enclosing a mass $M$ has a mass-to-flux ratio $M/\phi=\pi
c_s^2/(B_0 G^2 M)$.  With the values of $B_0$ listed above, and for a
region enclosing a mass $M=1$~$M_\odot$, this corresponds to a
mass-to-flux ratio, in units of the critical value, of $1.0$, $2.0$,
$3.0$, $3.6$ and $5.1$.

For each choice of $B_0$ and $\tau$ we ran 70 cases corresponding to 10
different values of the position angle $\phi$ and 7 values of the
inclination angle $\omega$. We considered both an isothermal
source and a radial temperature profile (see
Section~\ref{assumptions}). For each of the 2100 maps generated, the
model Stokes $Q$ and $U$ were compared with the observed values, and the
sum of individual $\chi^2$ was evaluated (see the radial temperature
profile results in Fig.~\ref{fig_galli93T-chi}).  The results show that
the best fit to the data is given by the models with the highest values
of the initial magnetic field $B_0$. In all cases, intermediate values of
$\phi$ and $\omega$ are selected.  Note also that for smaller values of
$B_0$, the best fit is achieved for lower values of $\omega$. This is due
to a zooming effect: a larger $B_0$ implies a smaller $r_0$.  For
large $B_0$ smaller angular distances mean larger radii, where the
magnetic field configuration of the outermost parts of the model are
naturally pinched in an edge-on view. On the other hand, for small
$B_0$ (large $r_0$), the innermost region the magnetic field tend to be
radial, and a larger inclination angle combined with the line-of-sight
emission integration is needed to produce the pinched morphology.

The fit is not sensitive to the value of the non-dimensional time
$\tau$. Thus, the time elapsed since the onset of collapse is not well
constrained by these models.  For example, for models with
$B_0=0.29$~mG (third column in Fig.~\ref{fig_galli93T-chi}) the time
corresponding to the three values of $\tau$ is $8.7\times 10^3$~yr,
$1.5\times 10^4$~yr, and $2.0\times 10^4$~yr, whereas for models with
$B_0=0.17$~mG (fifth column), the time range corresponding to the three
values of $\tau$ is 1.5--$3.5\times 10^4$~yr.

Fig.~\ref{fig_best-noT} and Fig.~\ref{fig_best-T} show the predicted
Stokes $I$, $Q$, $U$ maps and the Stokes $I$ residuals (second row)
compared to the observed maps (first row) for this model, for the case
with uniform temperature and temperature gradient, respectively.  Both
realizations shown have $B_0=0.43$~mG, $\tau=0.7$ (corresponding to
$t=1.4\times 10^4$~yr after the onset of collapse), $\phi=50^\circ$ and
$\omega=45^\circ$.  The derived physical parameters are shown in
Tables~\ref{tab_models-noT} and \ref{tab_models-T} for the case with
uniform temperature and temperature gradient, respectively.  The radial
temperature profile realizations show a better match to the
observational maps for all the Stokes maps, as well as physical
parameters closer to the observed toward \src\ . The general morphology
of the magnetic field could be reproduced, as well as the
depolarization toward the source axis and the double peak in polarized
emission.

\subsection{\citet{allen03a,allen03b}\label{ssec_allen03}}

\begin{figure}
\centering
\includegraphics[width=10cm,angle=0]{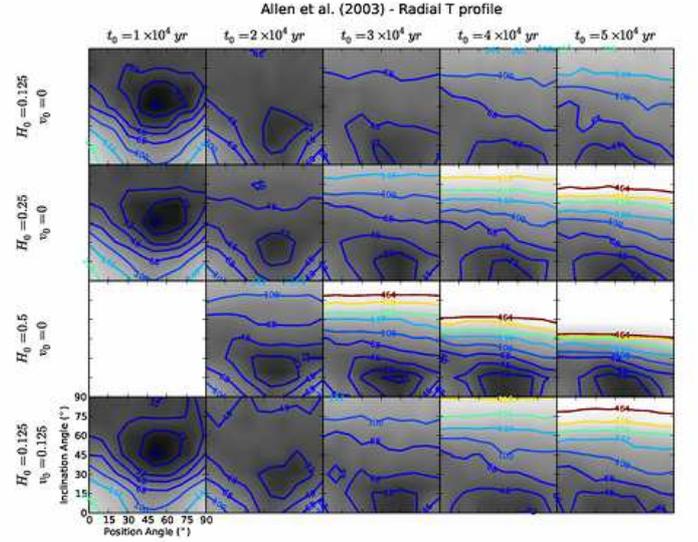}		
\caption{Same as Fig.~\ref{fig_galli93T-chi} for the
\citet{allen03a,allen03b} model with radial temperature profile.}
\label{fig_allen03T-chi}
\end{figure}

For the models of \citet{allen03a,allen03b} we selected 3 non-rotating
models with different values of the parameter $H_0$ defining the
mass-to-flux ratio of the initial state ($H_0=0.125$, 0.25 and 0.5,
corresponding to a mass-to-flux ratio in units of the critical value of
8.38, 4.51, and 2.66), and one rotating model, with $H_0=0.125$ and
uniform rotation velocity $v_0=0.125$ (in units of the sound speed).
For each case, we considered 5 evolutionary times, $t=10^4$~yr,
$2\times 10^4$~yr, $3\times 10^4$~yr, $4\times 10^4$~yr and $5\times
10^4$~yr.  As before, the position angles $\phi$ and $\omega$ were
varied over a grid of $10\times 7$ values, in both uniform and radial
temperature profiles, generating a total of 2660 maps (due to numerical
problems it was impossible to simulate the case $H_0=0.5$ with
$t=10^4$~yr).  The resulting $\chi^2$ of the comparison with the
observed Stokes $Q$ and $U$ maps for the radial temperature profile
cases is shown in Fig.~\ref{fig_allen03T-chi}.

It is evident from the figure that better fits are obtained with lower
values of the time elapsed since the onset of collapse, a few
$10^4$~yr.  The results are not very sensitive to the mass-to-flux
ratio nor to the rotation of the initial configuration. There is a
clear degeneracy between time and inclination angle with respect to the
plane of the sky:  a more concentrated field (a more pinched hourglass)
can be obtained by letting the model evolve, or by a larger inclination
of the magnetic field axis. For this reason, the region of minimum
$\chi^2$ moves towards lower values of $\omega$ at later times.

Fig.~\ref{fig_best-noT} and Fig.~\ref{fig_best-T} show the predicted
Stokes $I$, $Q$, $U$ maps and the Stokes $I$ residuals (third row)
compared to the observed maps (first row) for this model, for the case
with uniform temperature and temperature gradient, respectively.  Both
realizations shown have $H_0=0.125$, $v_0=0$, $t=10^4$~yr,
$\phi=50^\circ$ and $\omega=45^\circ$. The derived physical parameters
are shown in Tables~\ref{tab_models-noT} and \ref{tab_models-T} for the
case with uniform temperature and temperature gradient, respectively.
As for the \citet{galli93a,galli93b} models, radial temperature profile
realizations show a better match. For this model, the magnetic field
morphology, source axis depolarization, and double peak in polarized
emission could be reproduced like with the \citet{galli93a,galli93b}
models, although the polarized intensity derived was smaller in this
case.

\subsection{\citet{shu06}\label{ssec_shu06}}

\begin{figure}
\centering
\includegraphics[width=9cm,angle=0]{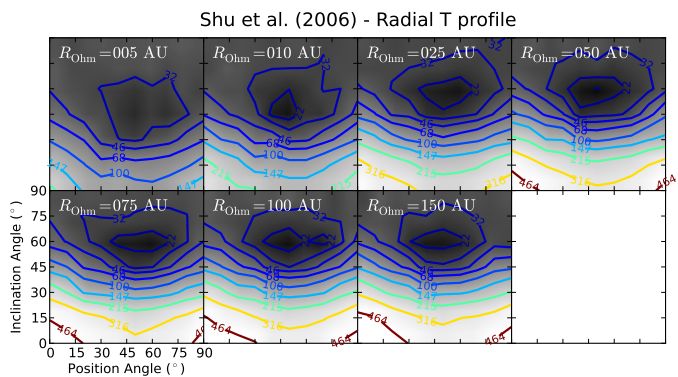}		
\caption{Same as Fig.~\ref{fig_galli93T-chi} for the
\citet{shu06} model with radial temperature profile.}
\label{fig_shu06T-chi}
\end{figure}

To test this model, we varied the Ohm radius from 5~AU to 150~AU. This
parameter controls the size of the region where magnetic dissipation
takes place and the magnetic field lines are almost straight. A total of
980 maps were generated, for 7 values of the Ohm radius, $10\times7$
values of the position and inclination angles, and both isothermal and
radial temperature profiles. Fig.~\ref{fig_shu06T-chi} shows the $\chi^2$
of the comparison with the observed $Q$ and $U$ maps for the radial
temperature profile case. The best-fit models tend to have $r_{\rm Ohm}$
in the range 10--100~AU, as also found previously
\citep{goncalves08}.

Fig.~\ref{fig_best-noT} and Fig.~\ref{fig_best-T} show the predicted
Stokes $I$, $Q$, $U$ maps and the Stokes $I$ residuals (fourth row)
compared to the observed maps (first row) for this model, for the case
with uniform temperature and temperature gradient, respectively. Both
realizations shown have $r_{\rm Ohm}=75$~AU,  $\phi=50^\circ$ and
$\omega=45^\circ$. The derived physical parameters are shown in
Tables~\ref{tab_models-noT} and \ref{tab_models-T} for the case with
uniform temperature and temperature gradient, respectively. As shown by
the figure, this model fails to reproduce the double-peaked
distribution of polarized intensity.   This feature is associated to
strongly concentrated, almost radial magnetic field lines in the
central region, at variance with the almost uniform field morphology
produced by magnetic dissipation. Another characteristic of this model
is the relatively high degree of polarization, not supported by the
observations.

\section{ALMA synthetic maps}
\label{sec_alma}

The ALMA resolution used (0\asec7$\times$0\asec4 $\sim$210$\times$120
AU$^2$) was chosen to have several projected cells of the simulations
($\sim$30~AU) inside each beam, allowing the best comparison possible
with the models avoiding resolution effects. The ALMA sensitivity and
{\it u,v} coverage is far much better than that of the SMA and, thus,
allows ({\it i}\/) to map with a much higher fidelity the polarized
emission, and ({\it ii}\/) to detect emission from a larger and fainter
region.

\begin{figure*} 
\centering 
\includegraphics[width=13.5cm,angle=0]{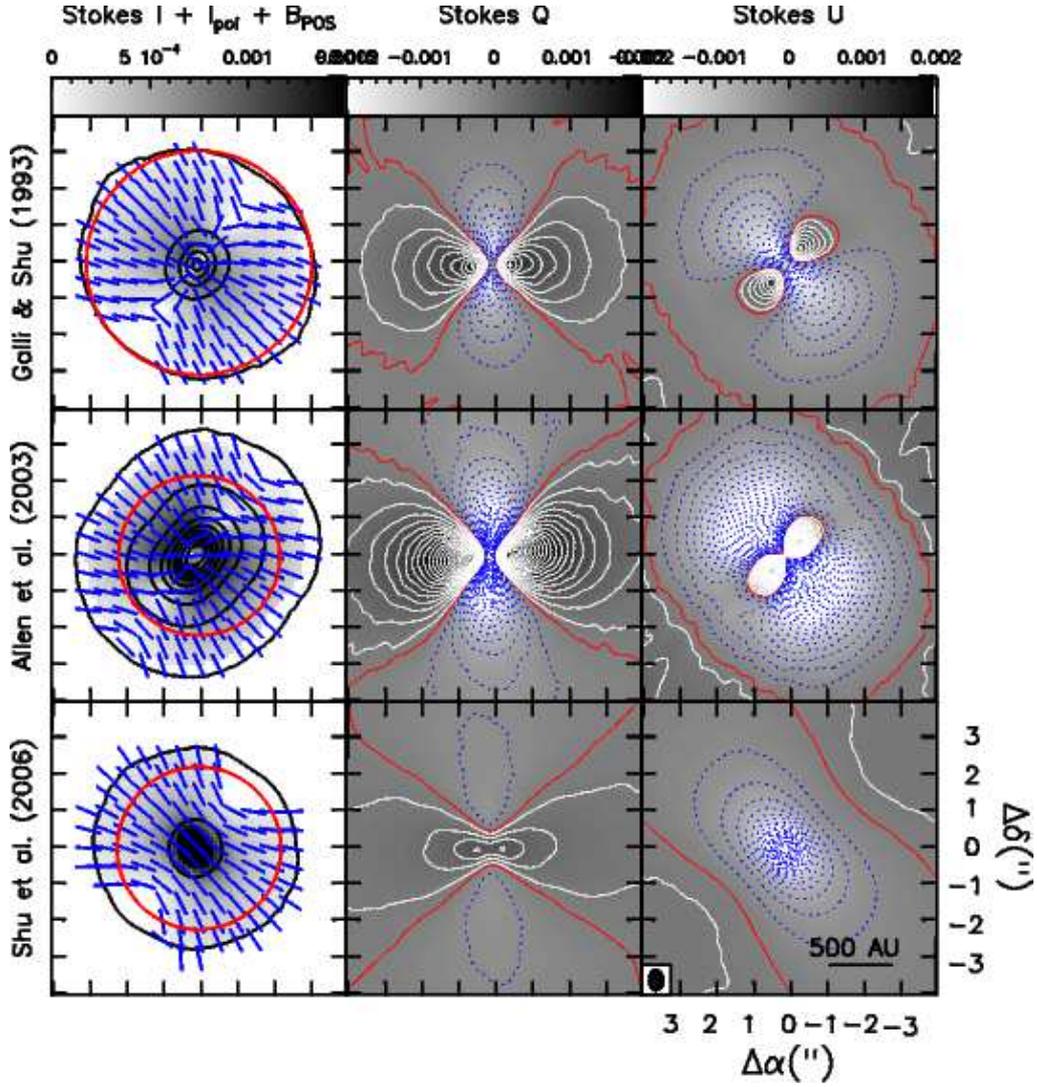}
\caption{ Same as Fig.~\ref{fig_best-T} for ALMA configuration~09.
{\it Contours}: Contours for the Stokes~$I$ maps (left panels) depict
emission levels from 5~$\sigma$ up to the maximum value in steps of
50~$\sigma$, where $\sigma$=0.2~mJy~beam$^{-1}$. Positive contours for
the Stokes~$Q$ and $U$ maps depict levels from 5~$\sigma$ up to the
maximum in steps of 10~$\sigma$, where $\sigma$=0.02~mJy~beam$^{-1}$.
Negative contours follow the same rule. {\it Red circles}: For the
\citet{galli93a,galli93b} and \citet{allen03a,allen03b} models depict
the loci of the front of the isothermal collapse wave ($r=c_s t$, see
Fig.~2 of \citealp{galli93a} and Fig.~6 of \citealp{allen03a}). For the
\citet{shu06} model it marks the 10~$r_{\rm Ohm}$ distance (see Fig.~4
of \citealp{shu06}).}
\label{fig_alma} 
\end{figure*}

Figure~\ref{fig_alma} shows the ALMA maps for the models shown in
Fig.~\ref{fig_best-T} using the radial temperature profile. The level
of detail of the convolved maps was very close to the original maps
thus an almost perfect morphology reconstruction is possible for all
the Stokes maps with the ALMA capabilities. Furthermore, the
combination of sensitivity and resolution achievable with ALMA makes
possible to extract usable information from the maps in a spatial range
$\sim 10$ times larger than the resolution. We derived large and
accurate polarization maps for all of the models. As
Fig.~\ref{fig_alma} states, it will be possible with ALMA to reach
resolution and detail levels which will allow to differentiate among
different models, and to select those matching better the observations.
In order to make this result more evident we marked in the left-hand
side panel of each model a red circle depicting important distances
related to the models. In the case of \citet{galli93a,galli93b} and
\citet{allen03a,allen03b} red circles depict the loci of the isothermal
collapse wave ($r=c_st$) which can be compared directly to Figs.~2 and
6 of the original papers, respectively. For the \citet{shu06} model,
red circle depict the 10~$r_{\rm Ohm}$ distance, comparable with Fig.~4
of their paper. The high power of reproduction of ALMA encourages
polarization observation toward all of the sources as a detailed
modeling will be possible with the onset of this powerful instrument.

\section{Summary and conclusions}
\label{summary}

The new data used in this work allowed to obtain a much better {\it
u,v} coverage for the \src\ region than in previous works, and,
therefore, more reliable maps. In addition, the data added from
different telescope configurations provided larger baselines, to
resolve the compact components, as well as shorter baselines, to better
trace the extended envelope emission. This significant improvement
allowed to separate the embedded compact sources from the diffuse
envelope. A good {\it u,v} coverage was essential to perform reliable
comparisons with models.  To this goal, we developed a selection
method, the so-called $\chi^2$ method, with larger discriminating power
than in previous studies (e.g. \citep{goncalves08}).

The new data confirm that the source emission is optically thin, with
no absorption or scattering, as expected for sub-mm emission. The
opacity derived from the Stokes $I$ map of the envelope is negligible,
implying that the maps trace very deep into the source. As the
scattering appear to be negligible, the origin of the alignment of the
dust grains is expected to be due to the magnetic field.

Despite the complexity of the NGC~1333 star forming region, MHD models of single
star formation assuming quasi-static initial conditions and a uniform (or
nearly uniform) magnetic field show remarkable agreement with the observed
characteristics of IRAS~4A, like intensity, polarized intensity, and
polarization distribution. These facts suggests that the dust polarization
pattern resulting from the density and magnetic field distribution of
non-turbulent models may apply even in less idealized initial conditions than
normally assumed. Once the orientation angles are consistently determined, the
comparison of the data with models of magnetized collapse indicate that a strong
initial field is required ($B_0$ larger than a few tenths of mG, see
sect.~\ref{model_galli}), that the source is very young (a few $10^4$~yr, see
sects.~\ref{model_galli} and \ref{model_allen}), and that the scale where
magnetic dissipation occurs is below the resolution of current observations (see
sect.~\ref{model_shu}). However, with the current level of sensitivity and {\it
u,v} coverage it is not possible to clearly discriminate
among different collapse models. The
\citet{galli93a,galli93b} and \citet{allen03a,allen03b} models fit better than
the \citet{shu06} model, but no selection can be done between the former two
models.

In general, the models predict sources with a more centrally peaked
core and a larger less dense envelope with polarized emission less
concentrated. A more refined dust grain treatment could help in a more
realistic emission treatment.  Current SMA observations of IRAS~4A
clearly favor models with a temperature gradient. The total emission
maps derived from models with a temperature gradient show the right
peak value but larger fluxes, steeper profiles and more extended
morphologies than \src. On the other hand, they predict the right
fluxes and morphologies of polarized emission but lower peaks and
softer profiles.  Conversely, polarization maps obtained with uniform
temperature profiles are more extended than \src\ maps (see 
Section~\ref{res_temp}). Therefore,
the inclusion of a realistic temperature profile cannot be ignored in
modeling the sum-mm emission of low-mass protostars.

An important result from the simulations of protostellar enveloped
threaded with an hourglass magnetic field is the possibility of deriving the
orientation angles of real sources from polarization measurements.
Figs.~\ref{fig_rot_QU} and \ref{fig_rot_I} show that up to inclination 
angles of $\sim 60$\deg\ it
is possible to estimate both position and inclination angles from the
Stokes $Q$ and $U$ maps. For larger inclinations the magnetic field tends
to be radial and the Stokes maps do not show significant differences.
Therefore, Figs.~\ref{fig_rot_QU} and \ref{fig_rot_I} can be used as 
templates for future 
observations of the dust polarized emission toward star forming cores.

Another remarkable result is the good agreement in the {\it u,v}-plane of the
observed and synthetic visibility amplitudes obtained assuming a
temperature gradient, except for the shortest baselines ($\lesssim
20$~$k\lambda$). The observed deficit of emission in \src\ with respect
to the models suggest a sharper density decrease at scales of
$\sim 1500$~AU than predicted by theoretical models.

The ALMA simulations have shown the capability of this new instrument to
distinguish fine details even between models of the same family. The
methodology used in this work has proved to be a powerful tool to compare
observations directly to theoretical models in a consistent way and avoiding
instrumental effects. Future polarization measurements with the ALMA
will provide real power to select the best models to describe the structure
and evolution of low-mass cores and, consequently, to disentangle the medium
conditions and the physics ruling the process. Upcoming radiative transfer
codes like ARTIST \citep{padovani11} will facilitate this kind of studies.
ALMA data, together with powerful radiative transfer codes, may be used
together with the technique developed in this work to extract as much
information as possible from the data and to constrain the models.

\begin{acknowledgements}

PF is partially supported by MICINN grant FPU. PF and JMG are supported 
by MICINN
grant AYA2008-06189-C03. PF and JMG are also supported by AGAUR grant
2009SGR1172. 
The authors are grateful to
Shih-Ping Lai and Tao-Chung Ching for gently sharing their SMA data.
The authors also thank Jos\'e Gon{\c c}alves and Jongsoo Kim 
for useful discussions, and the anonymous referee for useful
comments.

\end{acknowledgements}


\end{document}